\documentclass[twocolumn,showpacs,preprintnumbers,amsmath,amssymb]{revtex4}

\usepackage{graphicx}
\usepackage{dcolumn}
\usepackage{bm}
\usepackage{multirow}

\begin{document}

\title{Domain Wall Pinning and Potential Landscapes Created by Constrictions and Protrusions in Ferromagnetic Nanowires}

\author{Doroth\'ee Petit}

\affiliation{Nanoscale Magnetics group,
Department of Physics, Blackett Laboratory, Imperial College London\\
Prince Consort Road, London SW7 2BW, United Kingdom}%

\author{Ana-Vanessa Jausovec}%
\affiliation{Nanoscale Magnetics group,
Department of Physics, Blackett Laboratory, Imperial College London\\
Prince Consort Road, London SW7 2BW, United Kingdom}%

\author{Dan Read}
\affiliation{Nanoscale Magnetics group,
Department of Physics, Blackett Laboratory, Imperial College London\\
Prince Consort Road, London SW7 2BW, United Kingdom}%

\author{Russell P. Cowburn}
\affiliation{Nanoscale Magnetics group,
Department of Physics, Blackett Laboratory, Imperial College London\\
Prince Consort Road, London SW7 2BW, United Kingdom}%

\begin{abstract}
The potential experienced by transverse domain walls (TDWs) in the vicinity of asymmetric constrictions or protrusions in thin Permalloy nanowires is probed using spatially resolved magneto-optical Kerr effect measurements. Both types of traps are found to act as pinning centers for DWs. The strength of pinning is found to depend on the trap type as well as on the chirality of the incoming DW; both types of traps are seen to act either as potential wells or potential barriers, also depending on the chirality of the DW. Micromagnetic simulations have been performed that are in good qualitative agreement with the experimental results. 
\end{abstract}

\pacs{75.60.Ch, 75.75.+a, 75.60.Jk}

\maketitle

The ability to control DWs in nanoscale magnetic structures has become the focus of intense research in the past few years, for multiple reasons. 
The switching of thermally stable spin electronic devices of ever reduced lateral dimensions poses considerable power dissipation problems \cite{Cowburn03-JAP-93-9310}; DW mediated switching is a promising way of extending the scaling of spintronic devices to much smaller dimensions, 
leading to fundamental studies of the spin transfer effect where an electrical current is used to displace DWs along magnetic tracks \cite{Grollier02-JAP-92-4825,Thomas06-Nat-443-197,Lim04-APL-84-2820,Vernier04-EurophysLett-65-526}. Novel magnetic DW logic \cite{Allwood02-Science-296-2003,Allwood04-JAP-95-8264} and memory devices \cite{Allwood05-Science-309-1688,Ross06-JAP-99-08S501} have recently been demonstrated using controlled DW movement in complex magnetic nanowire networks. Furthermore, the ability to control the structure of a domain wall through the geometrical dimensions of the magnetic wire allows the experimental study of fundamental physical properties of these different types of DWs \cite{Klaui04-APL-85-5637,Klaui03-PRL-90-097202,Backes07-APL-91-112502}. 
As the size of nanomagnets continues to decrease, the role of surface structure, whether natural roughness \cite{Nakatani03-Nat-Mat-2-521} or patterned edges \cite{Herrmann00-JAP-87-2994} becomes more important. Deliberately fabricated defects in ferromagnetic nanowires allow control of the position of DWs \cite{Faulkner04-JAP-95-6717,Himeno03-JAP-93-8430}, as well as their propagation direction \cite{Allwood04-APL-85-2848,Bryan07}, as they create changes in the energy landscape which increase the local propagation field. 
Although there have been several experimental studies reporting the ability of artificially created constrictions to pin DWs \cite{Klaui03-PRL-90-097202,Klaui04-PhysicaB-343-343,Klaui05-APL-87-102509}, and numerous spin transfer experiments currently use such defects to precisely locate and hold DWs within magnetic nanostructures \cite{Grollier02-JAP-92-4825,Lim04-APL-84-2820}, a precise understanding of how the local DW energy landscape is modified by constrictions is currently lacking.
In this paper we report the use of spatially resolved MOKE to study the switching properties of L-shaped Permalloy nanowires containing single inward and outward notches. These high sensitivity measurements of a single DW in a single nanowire require the ability to detect magnetization reversals of about $10^{-12}$ emu. The handedness of the incoming DW is found to have a great impact on the strength of pinning as well as on the type of potential disruption created by the notch.

\begin{figure}[h!]
\begin{center}
\includegraphics[width=0.45\textwidth]{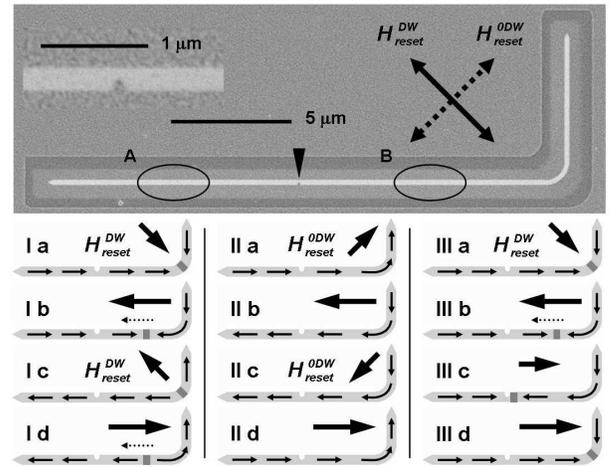}
\caption{Top: secondary electron image by FIB irradiation of a 200 nm wide Permalloy L-shape structure with an inward notch at its middle. Ellipses A and B show the positions of the measurements, the double arrows show the directions of $H^\textit{DW}_\textit{reset}$ and $H^\textit{0DW}_\textit{reset}$. Inset: high magnification image of the notch. Bottom: schematics illustrating the field sequences used to measure the switching properties of the magnetic structures. 
Ia to Id illustrate the creation and displacement of a DW, IIa to IId the nucleation field measurements, and IIIa to IIId the potential measurements. The large arrows indicate the direction of the external magnetic field, the narrow arrows the direction of the magnetization in the nanostructure, and the dotted arrows the direction of displacement of DWs.
}
\label{fig:LShape-Fields}
\end{center}
\end{figure}

200 nm wide L-shaped wires were milled using focused 30-keV Ga$^+$ ions from 7 nm thick, thermally evaporated Ni$_{80}$Fe$_{20}$ film on Silicon substrate. Nominally 100 nm wide, 100 nm deep inward and outward notches were patterned, either on the inner or on the outer side of the structures. A micrograph of an L-shaped nanostructure with an inward notch situated on the outer side of the horizontal section of the L-shape is shown at the top of Fig. \ref{fig:LShape-Fields}. 
\begin{figure}[t!]
\begin{center}
\includegraphics[width=0.45\textwidth]{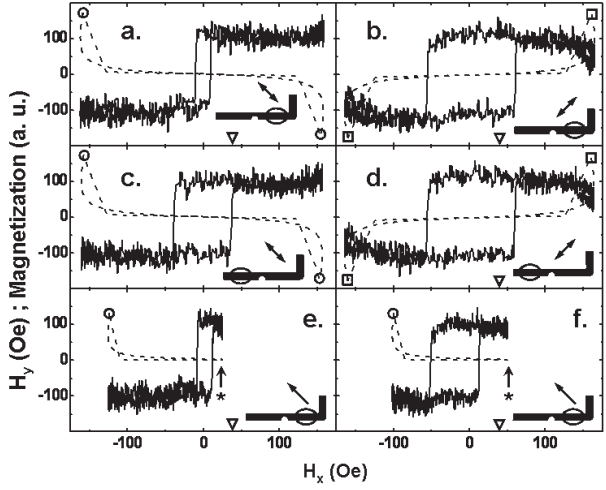}
\caption{MOKE-determined magnetization hysteresis loops (solid lines) and corresponding field sequences (dashed lines) used to measure a: the propagation field $H_P$; b, d: the nucleation field $H_N$; c: the transmission field $H_T$; e, f: the potential profile. The trap type-DW chirality is inward-up (type II: the trap is a constriction and the magnetization inside the core of the HtH DW created during the first half of the transmission field cycle points towards the trap). $\circ$ indicate $H^\textit{DW}_\textit{reset}$, $\square$ indicate $H^\textit{0DW}_\textit{reset}$, the arrows labelled with $\star$ show $H_{push}$, and $\triangledown$ show $H_T$ as measured in c. The schematics indicate on which part of the nanostructure the MOKE laser spot was focused and the direction of the reset pulse. }
\label{fig:AllLoops}
\end{center}
\end{figure}
Each structure was analyzed using a high sensitivity MOKE magnetometer. The $\sim 5\mu m$ diameter focused laser spot was placed on the horizontal section of the wire on either side of the notch. A quadrupole electromagnet was used to apply $H_x$ and $H_y$ magnetic fields at a frequency of 1Hz: $H_x$ (along the long part of the L-shape) is a sinusoidal field, and as it reaches its maximum (and minimum), an $H_y$ pulse is applied, producing a  45$^\circ$ reset field pulse of a few 100 ms duration.
Depending on the relative sign of $H_x$ and $H_y$, the resulting reset field pulse is either aligned with the dotted double arrow $H^\textit{0DW}_\textit{reset}$ of Fig. \ref{fig:LShape-Fields}, or with the solid one $H^\textit{DW}_\textit{reset}$; in the first case fully saturating the magnetization along the L-shape in order to have a defined initial magnetization state for nucleation field measurements ($H_N$), as shown Fig. \ref{fig:LShape-Fields} IIa and IIc, or creating a DW at the corner of the L-shape for propagation measurements in the latter case (Fig. \ref{fig:LShape-Fields} Ia and Ic). $H_y$ \textit{vs} $H_x$ graphs are shown Fig. \ref{fig:AllLoops}. In the case of propagation field measurements, two types of DWs are created for every field cycle, as illustrated at the bottom of Fig. \ref{fig:LShape-Fields}: at Ia, a downward head-to-head (HtH) DW is created at the corner of the L-shape; $H_y$ is then switched to zero, and the DW moves in the direction of the dotted arrow shown in Ib under the influence of $H_x$ only, going through the notch and annihilating at the end of the nanowire. An upward tail-to-tail (TtT) DW is then created during the second half of the field cycle (Ic) which subsequently propagates in the same direction through the notch (Id). 
The corresponding MOKE hysteresis loop measured between the corner of the L-shape and the notch (position B on Fig. \ref{fig:LShape-Fields}a) is shown Fig. \ref{fig:AllLoops}a in the case of an inward notch situated on the outer side of the L: a sharp transition is observed as $H_x$ reaches the propagation field $H_P$ necessary for the DW to move from the corner towards the notch, therefore reversing the magnetization in part B of the structure. When the laser spot is placed between the notch and the end of the nanowire 
\begin{figure}[t!]
\begin{center}
\includegraphics[width=0.45\textwidth]{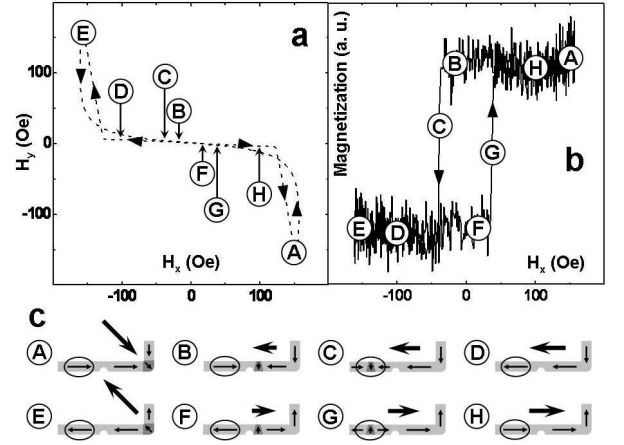}
\caption{Detail of the transmission field measurements shown Fig. \ref{fig:AllLoops}c, with simultaneous $H_y$ \textit{vs} $H_x$ curve (a), MOKE loop (b) and micromagnetic configuration induced in the nanostructure. The ellipse indicate where the laser spot is focused. The arrows on graphs indicate the sense of variation of the field as the sequence is applied. $H^\textit{DW}_\textit{reset}$ is applied (A), creating a HtH DW at the corner of the L and saturating the magnetisation in the horizontal part of the nanostructure in the positive direction. The field is then reversed and the DW is able to propagate towards the notch as soon as the field reaches the propagation field (B), reversing the part of the nanostructure situated between the corner and the notch. The DW then remains trapped there until the field reaches the transmission field, at which point the DW can further propagate (C), causing a sharp transition in the loop as the DW reverses the part of the structure situated between the trap and the free end, where the laser spot is focused. The DW then anihilates at the end of the nanowire and the magnetization in the horizontal part of the wire points in the negative direction (D). $-H^\textit{DW}_\textit{reset}$ is subsequently applied (E) at the start of the second half field cycle, creating a TtT DW at the corner by reversing the vertical part of the structure (the horizontal part already points in the negative direction) which propagates towards the notch as the field reverses and reaches the propagation field (F). The DW depins from the notch and reverses the part of the structure situated between the trap and the free end as the field reaches the transmission field, causing a sharp transition in the MOKE loop (G). The DW then also anihilates at the end of the nanowire (H).}
\label{fig:DetailAllLoops}
\end{center}
\end{figure}
(position A on Fig. \ref{fig:LShape-Fields}a), a sharp transition is observed as $H_x$ reaches the transmission field $H_T$ necessary for the DW to travel through the notch and reverse part A of the nanowire (see Fig. \ref{fig:AllLoops}c). Fig. \ref{fig:DetailAllLoops} shows the same transmission loop presented Fig. \ref{fig:AllLoops}c together with detailed schematics illustrating the evolution of the micromagnetic configuration induced in the nanostructure. $H_P$ and $H_T$ were measured at 10 Oe and 38 Oe. The loops measured at both positions using field sequence II are presented in Fig. \ref{fig:AllLoops} b and d. Both loops show the same $H_N$ of 58 Oe, ruling out the possibility that the transition shown Fig. \ref{fig:AllLoops}c is in fact due to nucleation of a new DW. 
Measurements in the vertical arm of the L-shape show that both arms of the structure switch simultaneously through the following mechanism: the horizontal part switches first, creating a DW at the corner, which immediately propagates upward and switches the vertical arm.
The loops measuring the transmission field (see Fig \ref{fig:DetailAllLoops} b for instance) are symmetric, showing that the upward HtH DW created during the first half field cycle and the downward TtT DW created during the second half field cycle have the same transmission properties, \textit{i. e.} the absolute direction of the magnetization in the core of the DW is not relevant as long as the rotationnal sense of the magnetization within the DW, or chirality, is preserved.
\begin{figure}[!t]
\begin{center}
\includegraphics[width=0.45\textwidth]{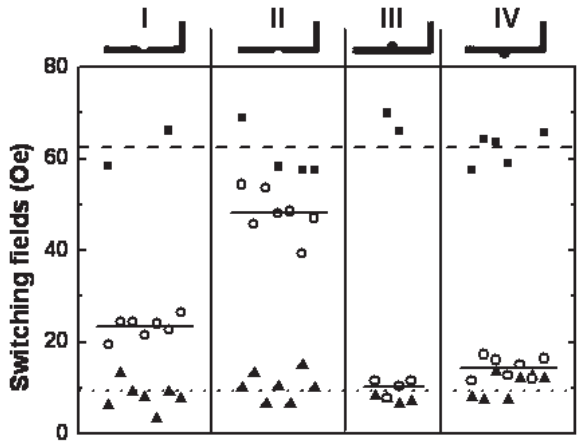}
\caption{Switching fields of nanowires containing inward and outward notches. Squares: nucleation field $H_N$; triangles: propagation field $H_P$; and open circles: transmission field $H_T$. The trap type - DW chirality of group I is inward-down (the trap is a constriction and the HtH DW created during the first half of the transmission field cycle points away from the side of the nanowire where the trap is), group II is inward-up (the trap is also a constriction but the HtH DW points towards the trap) , group III is outward-down (the trap is a protrusion and the HtH DW points away from it), and group IV is outward-up (protrusion and the HtH DW points towards it). The depining field from a trap depends on the type of the trap as well as on the chirality of the incoming DW.} 
\label{fig:Transmission}
\end{center}
\end{figure}
In order to probe the transmission properties of a DW of opposite chirality, one can either change the handedness of the corner of the nanowire and use $H^\textit{0DW}_\textit{reset}$ reset field pulse to create a DW, or keep the same corner chirality and the same reset field pulses and pattern the notch on the other side of the structure. We have chosen the second method. The magnetization inside the core of the HtH DW points either away from the notch (hereafter referred to as down) or towards the notch (up). 

$H_N$, $H_P$, and $H_T$ measured on 25 structures are presented Fig. \ref{fig:Transmission}. 
$H_P$ (triangles) and $H_N$ (squares) are fairly similar for all structures, at an average value of 9 Oe $\pm$ 3 (dotted line) and 63 Oe $\pm$ 5 (dashed line) respectively. On the other hand, there is a clear change in $H_T$ (open circles; the lines are group averages) as one changes either the type of the notch (inward or outward) and within the same notch type, as one changes the chirality of the DW. The average values of $H_T$ are 23 Oe $\pm$ 2 for group I, 48 Oe $\pm$ 5 for group II, 10 Oe $\pm$ 2 for group III, and 14 Oe $\pm$ 2 for group IV.

Having determined the strength of interaction between a DW and a range of artificial defects, we now consider the precise form of the potential modification.  We have developed a form of spectroscopy which uses the DW itself as a probe: 
a DW is created at the corner of the L-shaped structure using the same $H^\textit{DW}_\textit{reset}$ field pulse; $H_y$ is then set to zero and the DW is pushed towards the notch using an x-field which takes a maximum value $H_{push}$. The field $H_{pull}$ required to pull the DW back from the notch area to the corner is measured as a function of $H_{push}$. If the potential is a simple barrier, then it will be possible to pull the DW back with a field equal to $H_P$ as long as it has not gone over the barrier ($H_{push}<H_T$, see Fig. \ref{fig:AllLoops}e). However, a magnetic field equal to $H_T$ will be needed to pull the DW back out of a simple well. Mixed situations can occur where a well is surrounded by barriers or a barrier is surrounded by wells. In that case the different parts of the potential are probed by adjusting 
\begin{figure}[!t]
\begin{center}
\includegraphics[width=0.45\textwidth]{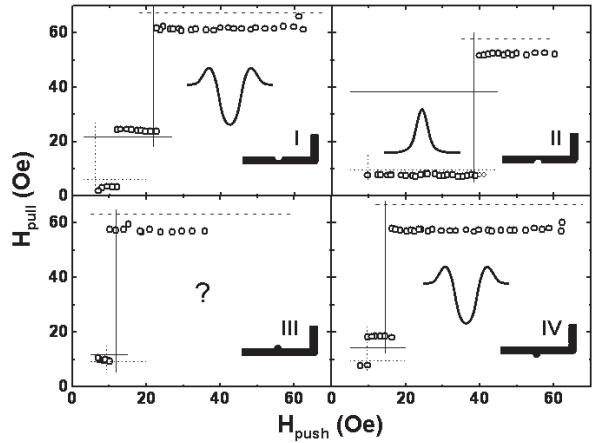}
\caption{Potential profile measurements: the field $H_{pull}$ necessary to pull a DW back from the trap area and into the corner is plotted as a function of the field $H_{push}$ used to initially push the DW towards the trap. 
Insets: schematics of the corresponding notch/DW configurations and of the type of potential landscape determined.}
\label{fig:Potential}
\end{center}
\end{figure}
the strength of $H_{push}$. The field sequence is described in Fig. \ref{fig:LShape-Fields}III and the results are shown Fig. \ref{fig:Potential}. If $H_{push}$ exceeds $H_T$ (horizontal and vertical solid lines), then the DW travels past the notch and annihilates at the end of the wire, and it is necessary to apply a field of the order of $H_N$ to reverse part B ($H_{push}>H_T$, see Fig. \ref{fig:AllLoops}f). Two different types of behavior are observed when $H_P<H_{push}<H_T$, \textit{i.e.} as the DW is pushed to the notch, but without traveling past it: group II defects act as a simple barrier, while group I and IV act as wells surrounded by small barriers (\textit{c. f.} the low $H_{push}<H_B=12$ Oe regime of Fig \ref{fig:Potential}I for instance, where $H_{pull}$ is of the order of $H_P$). No side wells were observed around the barrier in Fig. \ref{fig:Potential}II. The results obtained for group I agree with \cite{Klaui03-PRL-90-097202, Bedau07-JAP-101-09F509} where inward notches in the down configuration were found to attract TDWs. It is not clear experimentally what type of potential disruption is created by group III type notches, since $H_T$ and $H_P$ are so close for this configuration.

The film thickness and device lateral dimension studied here lie very close to the calculated stability limit between transverse and vortex type DWs \cite{McMichael97-IEEETransMag-33-4167,Nakatani05-JMMM-290-750}, but although both types of walls are energy minima for this geometry, \cite{Laufenberg06-APL-88-052507} shows that there is not enough thermal activation at room temperature for the TDW which is initially created to overcome the energy barrier which separates it from the vortex wall configuration. In order to understand the pinning mechanisms involved, we have reproduced the experiments using OOMMF \cite{OOMMF} simulations ($M_S=800 \times 10^3$ A/m, $A=13 \times 10^{-12}$ J/m, $3.5\times 3.5\times 7$ nm cell size). The simulations were performed quasistatically, with $\alpha$ set to 0.5 to speed up the calculations. In order to avoid any influence arising from the ends of the nanowire, the notch was placed in the middle of a $200$ nm wide and $4$ $\mu$m long wire with pointed ends. The magnetization configuration was initially forced with a TDW of the desired chirality next to the notch and both transmission and potential measurements were performed. 
\begin{figure}[!t]
\begin{center}
\includegraphics[width=0.45\textwidth]{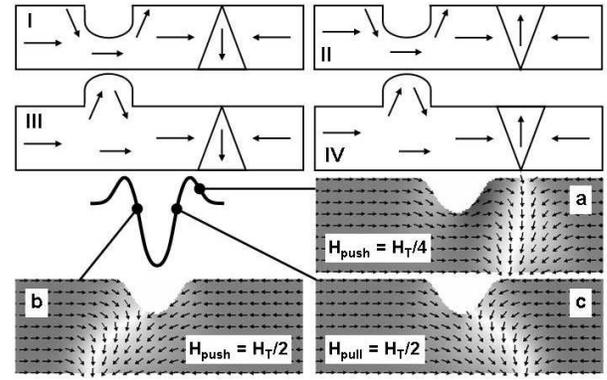}
\caption{I, II, III and IV: Schematics illustrating the micromagnetic configuration as a HtH DW approaches the notch in all four notch/DW configurations. a, b and c: OOMMF calculated configurations reproducing the potential measurements in the inward-down case (I). 
At (a), $H_{push}$ is not high enough to overcome the side barrier and the DW is pressed against the notch. As $H_{push}$ increases above $H_B$ (b), the DW falls into the main well and the DW is trapped in the notch. If $H_x$ is reversed at this point to try and pull the DW back (c), the same configuration as b is observed. It is only for $H_{push}=H_{pull}=H_T$ that the DW is able to leave the notch either way.
}
\label{fig:DWConfiguration}
\end{center}
\end{figure}
Fig. \ref{fig:DWConfiguration} I, II, III and IV schematically illustrate the DW/notch configuration obtained in the four cases studied here for the first half field cycle. As the downward HtH DW approaches the notch in the inward-down case (I), and for $H_{push}$ smaller than the characteristic field $H_B$, the calculations show that it first has to overcome the repulsion it experiences from the upward magnetization on the right side of the notch (side barrier, see Fig. \ref{fig:DWConfiguration}a). Once the right side of the notch has reversed and points downwards ($H_{push}>H_B$), then the narrow side of the V-shaped DW is maintained at the center of the notch, while the wide side keeps traveling towards the left as $H_x$ increases (central well, Fig. \ref{fig:DWConfiguration}b), extending the DW until $H_x$ reaches $H_T$ and the magnetization on the left side of the notch is finally reversed. $H_B$ and $H_T$ have been calculated at $22.5$ Oe and $82.5$ Oe. Both side barrier and central well have been observed experimentally. 
The opposite chirality (inward-up) is illustrated Fig. \ref{fig:DWConfiguration} II. In that configuration, the upward DW remains trapped on the right side of the notch where the local magnetization around the notch points in the same direction as the core of the DW (side well). In order to travel further, the DW has to overcome the energy barrier constituted by the downward magnetization on the left side of the notch (central barrier), which happens when $H_x$ reaches $H_T=227$ Oe. The DW has been found to depin from the side well for a reverse field of $72.5$ Oe. Only the central barrier is clearly observed experimentally (Fig. \ref{fig:Potential} II), although the experimental data are also compatible with the presence of a shallow side well. 
Fig. \ref{fig:DWConfiguration} III illustrates the outward-down case. In this configuration, the DW is first trapped inside the right side of the notch where the magnetization is also pointing downwards (side well, the depinning field from which has been calculated at $12.5$ Oe). In order to travel further, the DW needs to overcome the central energy barrier constituted by the upward magnetization inside the left part of the notch (central barrier). The narrow part of the V-shaped DW remains trapped inside the right side of the notch while the wide side keeps traveling towards the left as $H_x$ increases, until it reaches $H_T=62.5$ Oe and the left side of the notch (pointing upward) reverses. It was not possible experimentally to decide whether the structure of type III created a well or a barrier; the calculations show that it is a barrier with side wells.
Finally, the outward-up case is shown Fig. \ref{fig:DWConfiguration} IV. In this case, the DW has to initially overcome the side energy barrier constituted by the downward magnetization inside the right side of the notch. Once the first barrier is overcome ($H>37.5$ Oe) and the magnetization inside the notch is also pointing upwards, the wide part of the V-shaped downwards DW remains trapped, and the narrow part continues to travel towards the left until $H_x$ reaches $H_T=77.5$ Oe, at which point the DW can propagate further.
The experimental transmission fields are on average 80 percent lower than the calculated ones but no strict quantitative agreement is expected between calculations and experiments for the following reasons. The first one is the fact that simulations do not include the effect of temperature, whilst the experiments have been performed at room temperature. \cite{Himeno05} shows that DW depinning from a trap is a thermally activated process and reports a 40 to 60 percent decrease in the measured depinning fields as the temperature is increased from 4.2K to 300K. 
The second reason is due to the uncertainty on the exact geometry of the traps. Images of FIB fabricated notches suggest that their size spreads can be as large as 25 percent around the nominal value, with inward notches systematically larger than outward ones and slightly rounded corners. The width at half maximum of the simulated notches was taken at 28 pixels (98 nm), and the height was chosen so that all configurations allow the DW to transmit. Inward notches deeper than 22 pixels (77 nm) caused the DW in the up configuration to be unable to depin before a new DW nucleated on the other side, the heights/depths of the notches were therefore taken at 21 pixels (73.5 nm). Although the fabrication process caused protrusions to appear smaller than constrictions, we were interested in understanding the effect of the configuration rather than of the trap size, and therefore the same dimensions were used for both inward and outward notches. The exact geometries of the simulated notches are therefore different from the real ones, which contributes to the observed differences between the experimental and calculated pinning fields, as these are strongly affected by the trap dimensions \cite{Faulkner04-JAP-95-6717} and slope \cite{Allwood04-APL-85-2848}.
We have also performed all the calculations on structures with all lateral sizes divided by two. Although the transmission fields are much higher, the pinning potentials and relative pinning strengths are the same, showing the validity range of the DW pinning mechanism. 

Despite the quantitative difference, the qualitative agreement in the relative pinning strengths of the two different DW chiralities and the types of the potential disruptions is on the other hand excellent, and shows that the relative orientation between the DW core and the magnetization in the trap area plays a major role in determining the pinning potential: a well is observed if the DW core and the magnetization in the trap are parallel, a potential barrier if the DW core and the magnetization in the trap are antiparallel. It is not straightforward with this type of trap to clearly isolate the role of the DW/trap relative orientation and the role of the asymmetry in the DW shape, as this asymmetry is defined 
\begin{figure}[!t]
\begin{center}
\includegraphics[width=0.43\textwidth]{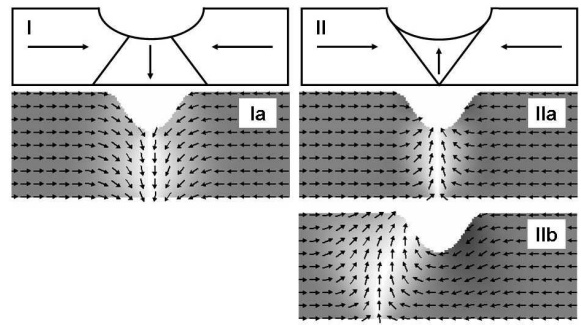}
\caption{Schematics (I, II) and OOMMF calculated (Ia, IIa, IIb) micromagnetic configuration obtained as TDWs of both chiralities are forced in the middle of a constriction (under zero field). In the case of configuration Ia, the trap edges are aligned with the magnetic moments inside the core of the DW (the trap edges are perpendicular to the V-shaped line shown in I). Configuration I (Ia) is stable, and the DW experiences a potential well. 
On the contrary, in the case of configuration IIa the trap edges are perpendicular to the magnetic moments in the core of the DW (the trap edges are parallel to the V-shaped line shown in II), leading to the creation of surface magnetic charges along the trap edge. Configuration IIa is therefore unstable; the DW experiences a potential barrier and spontaneously moves to the side of the trap where it can align its core magnetic moments with the trap edge (IIb).
}
\label{fig:StabIn}
\end{center}
\end{figure}
unilaterally by the charge and the chirality of the DW and because of the complex micromagnetic configuration in the traps. 
However, our present results suggest that the effect of the DW shape asymmetry has first of all to do with the fact that the DW charge is asymmetrically distributed within the DW: a TDW is a transversely oriented domain bounded on two sides by a V configuration of 45 Neel walls \cite{McMichael97-IEEETransMag-33-4167}, i. e. the magnetic moments along the V line are perpendicular to the V line. If the trap has its edges parallel to this line (configurations II and III), then placing the DW at the centre of the trap will create surface magnetic charges at the trap edges, and therefore be very costly energetically (barrier). On the contrary, if the defect has its edges perpendicular to the V lines of the DW (configurations I and IV), then placing the DW at the centre of the trap will reduce the demagnetizing energy (well), as such a defect has its edges aligned with the magnetic moments inside the DW. Fig. \ref{fig:StabIn} illustrates the point in the inward case:  
the main potential disruption created by an inward notch and as experienced by a DW in the downward configuration (I) is a well ultimately because it is energetically favorable for such a DW to sit in the middle of the notch. If it was energetically favorable for a DW of the opposite chirality (II) to sit at the centre of the same constriction, then this DW would be able, under a high enough magnetic field, to reverse the magnetization at the constriction so that it is aligned with its core magnetization, and the DW would experience a well preceded by a smaller well. It is not the case though. We can estimate the energy of the DW in this case by artificially forcing this unstable configuration in the simulations (see Fig. \ref{fig:StabIn} IIa): under zero field the energy of the DW is 40 percent higher than when the DW sits at the side of the trap (Fig. \ref{fig:StabIn} IIb).

In conclusion, we have performed MOKE measurements on 7 nm thick, 200 nm wide Permalloy L-shaped nanowires in order to study the pinning of TDWs by asymmetric constrictions and protrusions.
Both types of defects effectively increase the local propagation field. The chirality of the DW with respect to the notch has been controlled by patterning the latter either on the outer or on the inner part of the structure, and we have shown that it influences the strength of pinning as well as the pinning mechanism itself: both types of defects act either as potential wells or barriers depending on the chirality of the incoming DW. The main part of the potential profile ultimately depends on whether or not it is energetically favorable for a DW of a given chirality to sit in the middle of the notch, and whether or not the TDW is able, under a high enough external field, to switch the magnetization at the centre of the trap to align it with its core. Attractive potential wells are observed in the positive cases, repulsive potential barriers in the negative cases. Side barriers are observed around both potential wells, which reflect the initial antiparallel alignment between the DW core and the magnetization around the trap. The side well expected around the observed barrier (from the initial parallel alignment between the DW core and the magnetization around the trap) could not be detected experimentally, although the data do not rule out its presence.
Micromagnetic simulations qualitatively agree with the experimental results.

This work was supported by the European Community under the Sixth Framework Programmes MAGLOG (No. 510993) and SPINSWITCH (MRTN-CT-2006-035327).

\end{document}